\documentclass[superscriptaddress,twocolumn,prl,longbibliography]{revtex4-1}
\renewcommand{\vec}[1]{\mathbf{#1}}
\usepackage{amsfonts}
\usepackage{amsmath}    
\usepackage[normalem]{ulem}
\usepackage{bm}
\usepackage{graphicx}
\usepackage{titlesec}
\usepackage{tabularx}
\usepackage{color}
\usepackage{bbold}
\usepackage{upgreek,ulem,textcomp}
\usepackage{bbold}

\usepackage[colorlinks=true,linkcolor = black,citecolor = black]{hyperref}
\usepackage{hypernat}

\usepackage{amsfonts}
\usepackage{amsmath}    
\usepackage[normalem]{ulem}
\usepackage{bm}
\usepackage{graphicx}
\usepackage{titlesec}
\usepackage{tabularx}
\usepackage{color}
\usepackage{bbold}
\usepackage{upgreek,ulem,textcomp}
\usepackage{bbold}
\usepackage{comment}
\usepackage[british]{babel}
\usepackage{csquotes}
\usepackage{etoolbox}

\newcommand{\ket}[1]{\vert#1\rangle}

\newcommand{\Rmnum}[1]{\expandafter\@slowromancap\romannumeral #1@}

\begin{abstract}
Broadband spin-photon interfaces for long-lived storage of photonic quantum states are key elements for quantum information technologies. Yet, reliable operation of such memories in the quantum regime is challenging due to photonic noise arising from technical and/or fundamental limitations in the storage-and-recall processes controlled by strong electromagnetic fields. Here, we experimentally implement a single-photon-level spin-wave memory in a laser-cooled Rubidium gas, based on the recently proposed Autler-Townes splitting (ATS) protocol. We demonstrate storage of 20-ns-long laser pulses, each containing an average of 0.1 photons, for 200~ns with an efficiency of $12.5\%$ and signal-to-noise ratio above 30. Notably, the robustness of  ATS spin-wave memory against motional dephasing allows for an all-spatial filtering of the control-field noise, yielding an ultra-low unconditional noise probability of $3.3\times10^{-4}$, without the complexity of  spectral filtering. These results highlight that broadband ATS memory in ultracold atoms is a preeminent option for storing quantum light.
\end{abstract}

\begin{document}
\title{Single-photon-level light storage in cold atoms using the Autler-Townes splitting protocol}

\author{Erhan Saglamyurek}
\affiliation{Department of Physics, University of Alberta, Edmonton AB T6G 2E1, Canada}
\author{Taras Hrushevskyi}
\affiliation{Department of Physics, University of Alberta, Edmonton AB T6G 2E1, Canada}
\author{Logan Cooke}
\affiliation{Department of Physics, University of Alberta, Edmonton AB T6G 2E1, Canada}
\author{Anindya Rastogi}
\affiliation{Department of Physics, University of Alberta, Edmonton AB T6G 2E1, Canada}
\author{Lindsay J. LeBlanc}
\email{Corresponding authors: lindsay.leblanc@ualberta.ca, saglamyu@ualberta.ca }
\affiliation{Department of Physics, University of Alberta, Edmonton AB T6G 2E1, Canada}
\affiliation{Canadian Institute for Advanced Research, Toronto, ON, Canada}

\maketitle

For large-scale quantum networks to become practical, storage and on-demand recall of photonic quantum states must be available at timescales of up to several milliseconds~\cite{Lvovsky2009,Heshami2016a}. Interfacing non-classical light with these memories is necessary, but has proven to be difficult for two reasons: the substantial mismatch between the inherently large bandwidth of quantum light (from most popular single-photon sources) and the narrow acceptance bandwidth of well-studied atomic memories, and the unfaithful storage and recall processes due to photonic noise introduced by memory itself, which may degrade or fully destroy quantum nature of the stored light. This noise is particularly problematic with on-demand memories that require control electromagnetic fields, and is typically much more detrimental for broadband implementations~\cite{Lauk2013,Bustard2016,Ma2018}.

A promising approach to noise-free broadband memory is a family of  photon-echo-based protocols that feature inherently fast (non-adiabatic) memory operation~\cite{Tittel2009}. The Controlled Reversible Inhomogeneous Broadening (CRIB)~\cite{Moiseev2001,Kraus2006} and Gradient Echo Memory (GEM)~\cite{Alexander2006,Hetet:08} are widely studied protocols that rely upon the absorption of light via artificially broadened spectral features controlled by external electric or magnetic field gradients. However, implementing a broadband CRIB or GEM memory is technically challenging due to the infeasibility of large field gradients with rapid switching times. The atomic frequency comb (AFC) technique offers a solution to this limitation, as this approach relies on tailoring a comb-shaped spectral feature for light absorption without needing controlled broadening~\cite{Afzelius2009}. To this end, broadband AFC quantum memories have been successfully demonstrated for high-fidelity storage of entangled photons in the GHz-bandwidth regime using ensemble of two-level rare-earth (RE) ions in solids~\cite{Saglamyurek2011,Clausen2011a,Saglamyurek2015,Tiranov:15}. But, intrinsically short and pre-programmed storage times in these memories restrict their use to specific applications \cite{Sinclair2014,Saglamyurek2014}. The full AFC protocol in three-level systems can feature both long-lived storage and on-demand recall through collective spin excitations of atoms (spin-wave memory)~\cite{Afzelius2010}. However, well-known spin-wave compatible RE ions offer memory bandwidths of only a few MHz due to the small frequency spacing between the spin sub-levels, thereby hindering the protocol's suitability for a broadband memory~\cite{PhysRevA.88.022324,Gundogan2015}.

Another important avenue for broadband memory is the off-resonant Raman protocol, which features an all-optically controlled spin-wave memory~\cite{Gorshkov2007b,Nunn2007}. As this scheme relies on ``virtual'' absorption of light  with far-off resonant coupling, it can be used for light storage in atomic media with inhomonegeously broadened lines. Broadband Raman memories have been implemented in warm atomic ensembles which exhibit Doppler-broadening~\cite{Reim2010}. The operation of these memories in the quantum domain, however, has proven difficult due to large four-wave mixing noise, which cannot be eliminated using standard filtering techniques~\cite{Michelberger2015, Nunn2017}.  Noiseless broadband Raman memories have been achieved in diamond  (via storage on phononic transitions)~\cite{MacLean2015} and in ladder-type three level systems (via storage on optical transitions)~\cite{PhysRevA.97.042316, Finkelstein2018}, at the expense of losing the long-lived storage capability that comes with spin storage levels. Laser-cooled atoms provide a viable solution for a broadband spin-wave Raman memory, as demonstrated with the storage of 7-ns-long non-classical light pulses~\cite{Ding2015}. However, efficient storage of sub-ns pulses (GHz bandwidths) in these systems is technically very demanding in terms  optical depth and coupling-field power,  due to the inherent adiabatic operation of the Raman protocol, which exhibits unfavorable bandwidth scaling compared to fast memory protocols~\cite{Gorshkov2007f, Gorshkov2007b}.

\begin{figure*}
\begin{center}
\includegraphics[width = 178mm]{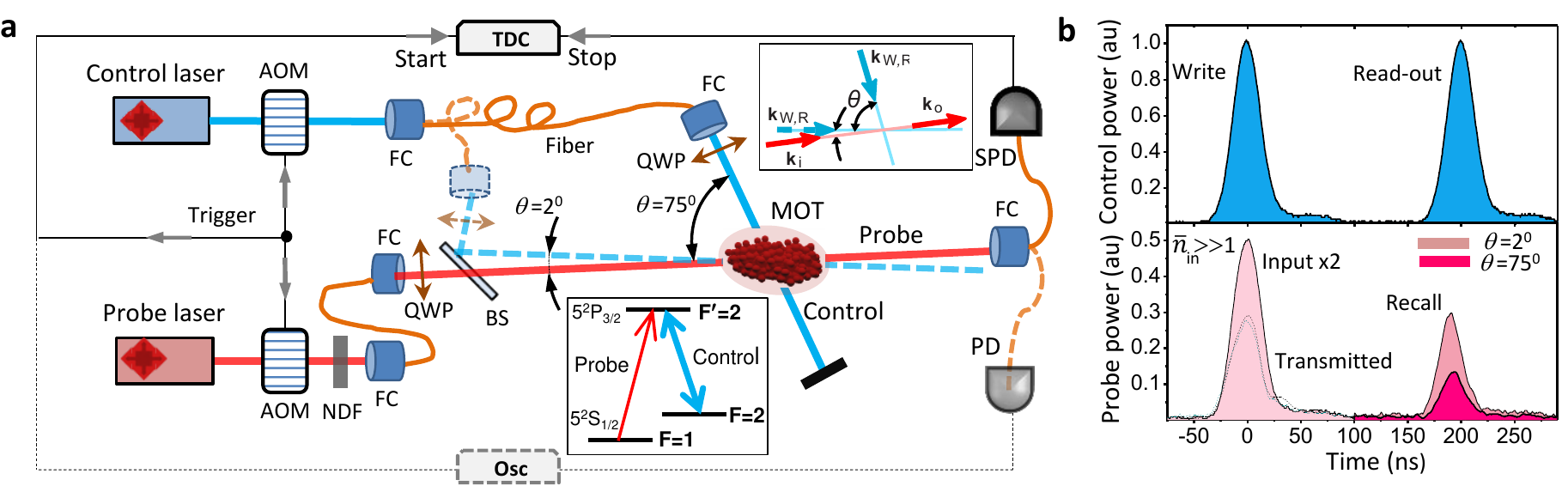}
\label{fig:setup}
\caption{\textbf{Experimental setup for implementing ATS memory:} \textbf{a.}~Control (blue) and probe (red) beams are derived from two independent continous-wave lasers and then shaped into short pulses using acousto-optic-modulators (AOM). After an adjustable attenuation of the probe beam with neutral density filters (NDF), both beams are coupled into single-mode fibers (FC), and decoupled back to free-space on a separate bench where magneto-optic-trap (MOT) apparatus is located. Following the polarization control using quarter-wave plates (QWP), the beams are overlapped in the atomic cloud (released from MOT) with a separation angle ($\theta)$ of either $2^{\circ}$ via a beam splitter (BS) (dashed traces) or $75^{\circ}$ (solid). After coupling into a fiber, the output probe is detected using either a standard photo-diode (PD) for $\overline{n}_{\rm in}\gg1$ (in dashed line) or a single-photon detector (SPD) for $\overline{n}_{\rm in}\leq1$. The arrival times of the detected signals (including directly transmitted probe) are recorded either on an oscilloscope (Osc) or time-to-digital convertor (TDC), respectively, triggered by a function generator (not displayed). \textbf{b}. 30-ns-long probe pulses with $\overline{n}_{\rm in}\gg1$ are stored and recalled in ATS memory (lower panel) at the $2^{\circ}$ (light) or $75^{\circ}$ (dark) separation angles using write and read-out control pulses (upper panel). Each trace is normalized to its maximum.}
\label{fig:setup}
\end{center}
\end{figure*}
The recently proposed Autler-Townes splitting (ATS) quantum memory protocol has great potential for overcoming these intrinsic and technical limitations \cite{Saglamyurek2018a}. ATS memory combines the inherently fast storage of photon-echo techniques with  all-optically controlled spin-storage of Raman-type adiabatic memories. 
 
In this study, we  explore  advantages of the ATS protocol and present  an ultra-low noise cold-atom based ATS memory, which operates, optimally, at rates much faster than those of atomic motion-induced spin decoherence. Under these conditions, the control-field noise is nearly eliminated using  all-spatial filtering techniques. We demonstrate reliable memory operation for short laser pulses, each containing a mean  photon number as low as 0.1, which is sufficiently low for quantum information processing~\cite{PhysRevA.72.012326}. Furthermore, by exploiting the simple pulse-area based operation of the ATS protocol, we experimentally realize a temporal beam-splitting process at the single-photon level to demonstrate memory-based pulse manipulation capabilities.


In our experiments, we use a $\Lambda$-type three-level system within the $^{87}$Rb atom's electronic structure, comprised of two ground hyperfine levels ($\ket{F= 1}\equiv\ket{g}$ and $\ket{F= 2}\equiv\ket{s}$, energetically spaced by 6.83 GHz) and an excited level ($\ket{F^\prime = 2}\equiv\ket{e}$) on the 780 nm ``D2'' transition (inset of Fig.~\ref{fig:setup}a). The atoms are laser-cooled in a standard magneto-optic trap (as described in Ref.~\cite{Saglamyurek2018a}), followed by further sub-Doppler cooling, and finally optical pumping for populating the atoms in $\ket{g}$ level. The atomic cloud is probed for light storage/retrieval after releasing it for $6$ ms time-of-flight, which yields an optical depth of $d\approx10$.  

We implement the ATS protocol in this system for storage of weak probe laser pulses (resonant with $\ket{g}\rightarrow\ket{e}$ transition), containing an average number of photons between $\overline{n}_{\rm in}= 0.1$ and $4\times10^{6}$. In this protocol~\cite{Saglamyurek2018a}, a strong control field (coupled to the $\ket{s}\rightarrow\ket{e}$  transition) with the pulse-area of $2\pi$ dynamically splits the natural absorption line into two peaks, as per the Autler-Townes effect. Upon absorption of the probe pulses, optical coherence is transiently mapped between the ground ($\ket{g}$) and excited level ($\ket{e}$), and then efficiently transferred onto the ground levels ($\ket{g}$ and $\ket{s}$) as a  collective spin excitation for storage (writing stage, Fig.~\ref{fig:setup}b). Retrieval is accomplished after an adjustable time by applying a second control pulse with  pulse area of $2\pi$, which reconstructs the photonic signal, following a brief re-establishment of coherence between $\ket{g}$ and $\ket{e}$ (read-out stage). Memory efficiency can approach unity, with minimal demand on optical depth and control-field power when probe pulses are much shorter than the coherence decay time of the excited level. Compared to the adiabatic memory schemes, such as the off-resonant Raman and Electromagnetically Induced Transparency (EIT) protocols~\cite{Saglamyurek2018a, Rastogi2019a}, these relaxed requirements for broadband memory provide a great advantage for eliminating  control-field-related noise, which is an increasing function of optical depth and/or control intensity~\cite{Ma2018, Nunn2017, Lauk2013}. Furthermore, the inherently high-speed operation of ATS memory enables all-spatial filtering of the control noise under certain conditions, which are not accessible to adiabatic memory techniques, as described in the following.  

In our implementation, sketched in Fig.~\ref{fig:setup}a, the direction of the control beams (represented by wave vectors of $\vec{k_{\rm w}}$ for write pulse and $\vec{k_{\rm R}}$ for read-out pulse) relative to the input probe beam ($\vec{k_{\rm i}}$) determines the direction of the output (retrieved) probe ($\vec{k_{\rm o}}=\vec{k_{\rm i}}-\vec{k_{\rm W}}+\vec{k_{\rm R}}$).  The relative angles play a key role in  both the memory's efficiency and the amount of noise stemming from the control field~\cite{Surmacz2008}. In the experiments here, output probe pulses are retrieved in the same direction as input probe pulses (``forward recall'' with $\vec{k_{\rm i}}=\vec{k_{\rm o}}$), using co-propagating write and read-out control pulses ($\vec{k_{\rm W}}=\vec{k_{\rm R}}$) which, despite technical ease, limits the theoretical maximum memory efficiency to $54\%$ due to re-absorption~\cite{Saglamyurek2018a}. In this arrangement, we spatially separate the probe and control beams that overlap in the atomic cloud by introducing an angle $\theta$ between them. While this angle allows for substantial extinction of the control photons from the probe spatial mode before detection, it also leads to a spatial phase-grating for the stored spin-wave with a period of $\kappa=2\pi/|\Delta\vec{k}|$, where $\Delta\vec{k}=\vec{k_{\rm i}}-\vec{k_{\rm W}}$ is imposed by conservation of momentum (phase-matching condition)~\cite{Surmacz2008,Zhao2008}

To begin, as in many previous memory implementations with cold atoms, we set a small separation angle $\theta=2^{\circ}$ (dashed control-beam trace in Fig.~\ref{fig:setup}a), resulting in  control-field extinction of $\approx40$~dB, as well as $\kappa\approx23~\mu$m.  With this spin-wave periodicity, approximately  $100~\mu$s would have to elapse before our cold atoms (at a temperature of $50~\mu$K) would diffuse and thus ``erase'' the spin-wave grating during storage, due to motional decoherence. Moreover, for additional spatial filtering of the control field, we collect the probe beam using a single-mode fiber, which increases the isolation  up to $65-70$~dB.  

\begin{figure}
\begin{center}
\includegraphics[width = 84mm]{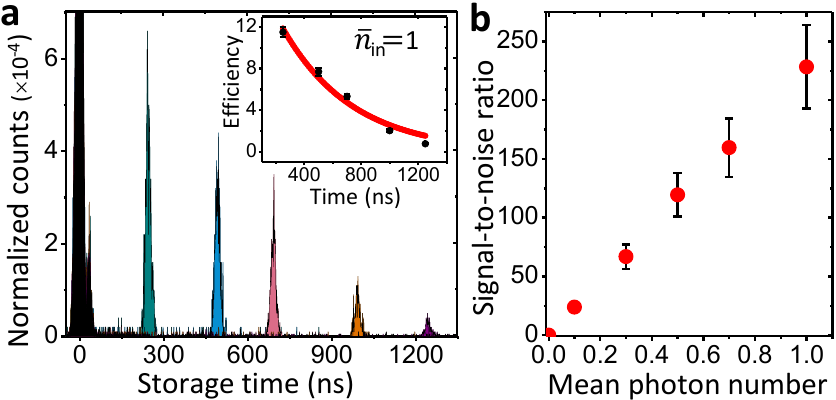}
\label{fig:single-photon}
\caption{\textbf{Single-photon-level ATS memory.} \textbf{a.} Six detection histograms (each normalized to the total number of storage-and-recall attempts $N=5\times10^4$ to $9\times10^5$), recorded for different storage times (between $250$ ns and $1250$ ns) to demonstrate on-demand memory using 30-ns-long input pulses with $\overline{n}_{\rm in}=1$. The inset shows the decay of memory efficiency, which is fitted to an exponential curve (solid line). \textbf{b.} Signal-to-noise ratio (SNR) vs.~mean photon number ($\overline{n}_{\rm in}$) for 30-ns-long probe pulses stored in ATS memory for 200 ns. The detection probabilities for recalled probe ($p_{\rm s}$ for a given $\overline{n}_{\rm in}\neq0$) and noise ($p_{\rm n}$ for $\overline{n}_{\rm in}=0$) are the ratios of the total detection counts (over $\Delta t=50$~ns centered around the recall time) to their respective $N=10^{5}$ to $9\times10^{5}$, depending on $\overline{n}_{\rm in}$.}
\label{fig:single-photon}
\end{center}
\end{figure}

In this first configuration, we assess the performance of our memory using probe pulses with large mean photon number ($\overline{n}_{\rm in}\approx4\times10^{6}$). We store 30-ns-long Gaussian pulses (at full-width-half maximum) and recall after 200 ns, using the write and read-out control pulses with the same temporal profile as the probe, and  peak power of $\approx20$~mW, which gives a pulse area of about $2\pi$ (Fig.~\ref{fig:setup}b). We measure the memory efficiency $\eta=23\%$, a threefold improvement upon our first demonstration~\cite{Saglamyurek2018a}, but still smaller than the theoretical maximum of $38\%$ for $d=10$ in the forward-recall configuration. This deviation from the theoretical maximum is mainly because of the magnetic-field induced spin-wave decoherence (as will be further discussed), and partly due to the spatially non-uniform control power that induces additional decoherence during the transient storage between the ground and excited level. Furthermore, we estimate the average number of the control photons leaking to the probe mode to be on the order of a few hundreds, which is still much smaller than the photon number contained in our probe pulses. While the influence of this leak is negligibly small in this large mean-photon-number regime, it is a major obstacle for memory operation at the single-photon level.

A standard technique to eliminate such noise is to use a single or a series of  spectral filters after recall, which can reject the control photons, detuned by $6.83$ GHz in our case, at the expense of additional loss and complexity. Instead of this approach, we follow the strategy of improving spatial filtering by increasing the separation angle between the control and probe beam from $2^{\circ}$ to $75^{\circ}$ (solid control-beam trace in Fig.~\ref{fig:setup}a.). To our detriment, such a wide separation angle also induces a large motional decoherence due to a small-periodicity  spin-wave grating ($\kappa\approx0.65~\mu$m), which would be washed out over a time scale of only $\approx1.5~\mu$s at a cloud temperature of $50~\mu$K. Nevertheless, the fast storage/retrieval  of the ATS protocol enables efficient memory operation at time scales much shorter than the motional decoherence time, whereas adiabatic memories that would require  probe pulses nearly a microsecond long in our limited-optical-depth experimental conditions (see Ref.~\cite{Rastogi2019a} for details).

With this trade-off between the large control-field filtering and limited storage times, we observe that the number of control photons leaking to the probe mode reduces from $\approx 300$ to the single photon-level (over 4 orders of magnitude extinction), whereas the memory efficiency decreases by only a factor of $\approx 2$ as compared to the small-angle setting for the same storage time (Fig.~\ref{fig:setup}b). We attribute this degradation of memory efficiency at large angles to the limited interaction region that arises from a spatial mismatch between the probe and control beams, having diameters comparable to the size of our atomic cloud.

After  substantially isolating the  probe beam from stray control photons, we now evaluate the operation of ATS memory at the single-photon level with time-resolved photon-counting measurements using a single-photon detector (SPD) and a time-to-digital converter (TDC). First, we demonstrate the storage and on-demand retrieval of the 30~ns-long probe pulses with $\overline{n}_{\rm in}=1$, as shown in Fig.~\ref{fig:single-photon}a. In this demonstration, the storage time is varied between 250~ns and 1250~ns by  changing the time difference between the write and read-out pulses. We observe that memory efficiency drops from $\eta=(11.5\pm0.5)\%$ to $(0.8\pm0.1)\%$ over this time interval, which we attribute to spin-wave decoherence (inset of Fig~\ref{fig:single-photon}a). The memory lifetime (at 1/$e$) is $490\pm60$~ns and is close to the one measured for the small-angle setting ($\approx650$~ns), indicating that spin decoherence due to non-zero ambient magnetic-fields dominates over the motional decoherence in our experiments.

 \begin{figure}
\begin{center}
\includegraphics[width = 84mm]{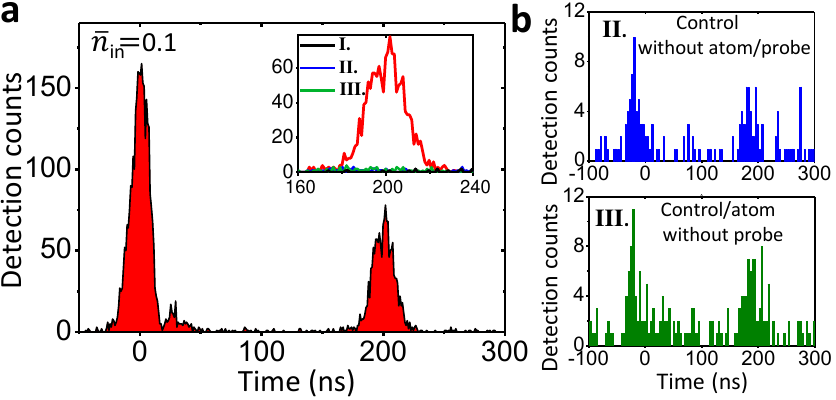}
\label{fig:noise}
\caption{\textbf{Noise characterisations:} \textbf{a.} Detection histogram for storage of 20-ns-long probe pulses for 200~ns with $\overline{n}_{\rm in}=0.1$ and $N=1.1\times10^{6}$, using  1~ns/bin. Detections at around zero-time are due to the non-absorbed (transmitted) part of input signal in the memory medium. The inset shows detections for recalled photons, with respect to noise that is measured in three configurations: (I) Probe without control/atoms, (II) Control without probe/atoms (III) Control/atoms without probe. \textbf{b.} Detection histograms with larger time-bins (4~ns/bin) for the noise measured in II (upper panel) and III (lower panel). }
\label{fig:noise}
\end{center}
\end{figure}

 Second, we lower the mean photon number of the input probe pulses below unity and characterise the signal-to-noise ratio ${\rm SNR}=(p_{\rm s}-p_{\rm n})/p_{\rm n}$ as a function of $\overline{n}_{\rm in}$ after 200 ns storage in memory, where $p_{\rm s}$ and $p_{\rm n}$ are independently measured detection probabilities for retrieved probe and noise (in the absence of probe) after $N$ trials, respectively   (Fig.~\ref{fig:single-photon}b). We observe an almost-linear dependence of SNR on $\overline{n}_{\rm in}$, and  measure a $\rm{SNR}=24\pm5$ for $\overline{n}_{\rm in}=0.1$,

 Third, we investigate the source and influence of the observed residual noise in a more demanding memory implementation: one with larger bandwidth (requiring larger control power) and lowest mean photon number at the memory input. For this purpose, we decrease the duration of the probe pulses from 30~ns to 20~ns, which is technically the shortest possible in our experimental setup. This, in turn, requires shortening the control pulses by the same amount, while increasing the control power to maintain the pulse area near $2\pi$. Under these conditions, we set $\overline{n}_{\rm in}=0.1$ for probe pulses, and store them in ATS memory for 200~ns with $\eta=(12.5\pm0.4)\%$, as shown in Fig~\ref{fig:noise}a. After $N=1.1\times10^{6}$ storage-and-recall attempts, we measure a total number of detection counts $N_{\rm s}=1357\pm40$ over a time-window of $\Delta t=30$~ns centered around the recall time. Next, we analyze noise contributions involved in the storage-and-recall process by determining the total detection counts in three complementary configurations for the same time-window and number of attempts, as depicted in the inset of Fig.~\ref{fig:noise}a. The first configuration establishes the background noise (due to ambient photons and detector dark count) in the absence of control field and atoms, yielding $N_{\rm 1}=5\pm2$ counts. In the second and third configurations, the write and the read-out pulses are launched in the absence of the probe field without atoms and with atoms, yielding $N_{\rm 2}=22\pm5$ and $N_{\rm 3}=36\pm6$ counts, respectively. Comparing these three results shows that the main noise component is due to photons from the scattered control field, while the background noise is insignificant. 
 
 For further confirmation, we analyze the detection histograms from the second and  third configurations using a larger bin size (4 ns/bin, instead of 1 ns/bin), as shown in Fig.~\ref{fig:noise}b. We observe two distinct peaks appearing with 200 ns separation, which corresponds to the time difference between the write and read-out pulses. Moreover, we determine that these peaks are shifted backward in time by about $24$ ns compared to the original arrival times of the control pulses. Relative to this timing, the number of the detected counts (within $\Delta t=30$ ns) is measured to be nearly the same for each peak without atoms and with atoms (${N_{\rm 2}}^{\rm W}=41\pm6$ and ${N_{\rm 3}}^{\rm W}=41\pm6$ for writing, and ${N_{\rm 2}}^{\rm R}=31\pm6$ and ${N_{\rm 3}}^{\rm R}=38\pm6$ for reading, respectively), conclusively showing that these noise counts are mainly due to the stray control-field photons in our setup, which could be  eliminated spectrally. This result also implies that there is no measurable four-wave mixing noise for these experimental conditions. 
 
 With these noise contributions (all included in the third configuration), we determine the unconditional-noise detection probability per pulse (excluding the total loss factor of $\eta_{\rm t}\approx0.1$ after memory for the worst-case scenario) to be ~$p_{\rm n}=N_{\rm 3}/N\times(1/\eta_{\rm t})=(3.3\pm0.5)\times10^{-4}$, which leads to $\rm{SNR}=37\pm6$ for the stored-and-recalled 20-ns-long probe pulses. Based on this value, the fidelity of our memory is estimated as $\mathcal{F}=1-1/\rm{SNR}=0.97\pm0.01$, if quantum states were stored. These results demonstrate that the ATS memory can be readily interfaced with non-classical light sources for high fidelity storage of photonic quantum states.  
 
  \begin{figure}
\begin{center}
\includegraphics[width = 84mm]{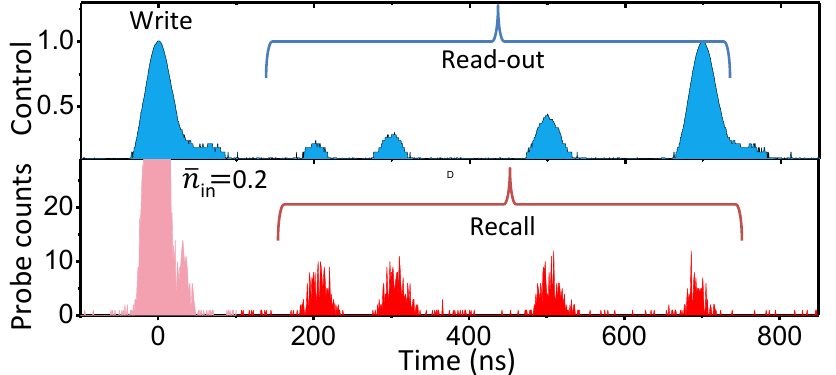}
\label{fig:split}
\caption{\textbf{Temporal beam splitting at the single-photon level}. Upper panel: A write control pulse and multiple read-out pulses, each of 30~ns duration, are used for retrieval of a single stored mode at different times with adjustable amplitudes. Control pulses, shown in units proportional to electric field/Rabi frequency, have pulse areas $\leq 2\pi$ depending on the fraction of spin coherence desired upon retrieval. These pulses are normalized to the height of the $2\pi$ write pulse. Lower panel: Detection histogram obtained after $N=4\times10^{5}$ attempts for storage and retrieval of 30-ns-long probe pulse with $\overline{n}_{\rm in}=0.2$, using the control-pulse sequence in the upper panel.}
\label{fig:split}
\end{center}
\end{figure}

 Finally, we demonstrate temporal beam splitting using ATS memory at the single-photon level (Fig~\ref{fig:split})~\cite{Saglamyurek2018a}. We store 30-ns-probe pulses with $\overline{n}_{\rm in}=0.2$ using a write pulse with pulse area of $2\pi$, as in the standard ATS scheme. In contrast, the retrieval is sequentially realized at different times using multiple read-out control pulses, each with a pulse area smaller than $2\pi$ (except the last one that retrieves all the remaining spin coherence). This process results in an output photon in a superposition of multiple temporal bins, as shown in the lower panel of Fig~\ref{fig:split}. The temporal beam splitting feature of ATS memory can be used for various applications in photonic quantum information, such as those that rely on time-bin qubit encoding.  
 
Beyond these proof-of-principle demonstrations, future implementations of ATS memory could benefit from the use of ultracold atoms in an optical-dipole trap~\cite{Chuu2008}. First of all, the storage time can be extended to the millisecond regime by using clock spin-states together with a proper magnetic field cancellation~\cite{Zhao2008a}, or even to the seconds-time scale~\cite{Dudin2013}, given the long hold times of atoms in optical-dipole traps. Particularly, a Bose-Einstein condensate (BEC) offers both long storage times and flexibility of all-spatial filtering simultaneously, as atomic diffusion in a BEC is virtually absent~\cite{PhysRevA.85.022318}. Second, the memory efficiency can reach near-unity in the backward recall configuration with an increase of the optical depth from the current value of $d\approx10$ to $d\approx100$, which is feasible with the inherently large densities of ultracold quantum gases~\cite{Vernaz-Gris18}. Third, the memory bandwidth can be extended from 20 MHz to the GHz regime, by forming the $\Lambda$ system in the D1 manifold of Rb, which provides near-GHz excited-level spacing, and by using fast electro-optic components for tailoring control pulses. With these moderate improvements, the ATS memory approach provides a practical solution to develop a high-performance quantum memory required for quantum networks.

In conclusion, we experimentally implemented the ATS protocol in a cold Rb gas for storage of single-photon level optical pulses as collective spin excitations. Ultra-low-noise memory operation is achieved for input photon numbers as low as 0.1 by eliminating the contamination of the control-field photons with all-spatial filtering techniques, at the expense of limited storage times. While the residual noise is dominated by the stray control field, the four-wave mixing noise is not a limiting effect in our experimental configuration. These results indicate that cold-atom based ATS memory can readily operate in the quantum regime for storage of non-classical states of light, which is our next goal towards the realization of a truly practical quantum memory.

\acknowledgments{We thank Dr. Khabat Heshami for useful discussions, and Dr. Alex Lvovsky for lending us a single-photon detector for our initial measurements. We appreciate generous technical support from Paul Davis and Greg Popowich.  We gratefully acknowledge funding from the Natural Science and Engineering Research Council of Canada (NSERC RGPIN-2014-06618), Canada Foundation for Innovation (CFI), Canada Research Chairs Program (CRC), Canadian Institute for Advanced Research (CIFAR), Alberta Innovates (AITF), and the University of Alberta.}

\textit{Note added in preparation}: Recently, S.A. Thomas et al.~\cite{Sarah2019} posted a result demonstrating  built-in noise suppression in an off-resonant Raman scheme using a new technique based on a specific detuning operation. 


\end{document}